\documentclass[prl,twocolumn,preprintnumbers,amsmath,amssymb]{revtex4}
\usepackage{graphicx}
\usepackage{latexsym}
\usepackage{amsmath}
\usepackage{graphics}
\usepackage{amssymb}
\usepackage{layout}
\usepackage{verbatim}
\usepackage{amsfonts,epsfig}
\usepackage{bbm}
\usepackage{comment}
\newcommand{\ket}[1]{\left|#1\right>}

\newcommand{\beq}{\begin{equation}}
\newcommand{\eeq}{\end{equation}}
\newcommand{\bea}{\begin{eqnarray}}
\newcommand{\eea}{\end{eqnarray}}
\newcommand{\nn}{\nonumber}

\usepackage{subfigure}



\begin{document}

\title{Exact Classification of Landau-Majorana-St\"uckelberg-Zener Resonances By Floquet Determinants}

\author{Sriram Ganeshan}
\author{Edwin Barnes}
\author{S. Das Sarma}
\affiliation{Condensed Matter Theory Center and Joint Quantum Institute, Department of Physics, University of Maryland, College Park, MD 20742, USA}

%%%%%%%%%%%%%
\begin{abstract}
Recent experiments have shown that Landau-Majorana-St\"uckelberg-Zener (LMSZ) interferometry is a powerful tool for demonstrating and exploiting quantum coherence not only in atomic systems but also in a variety of solid state quantum systems such as spins in quantum dots, superconducting qubits, and nitrogen vacancy centers in diamond. In this work, we propose and develop a general (and, in principle, {\it exact}) theoretical formalism to identify and characterize the interference resonances that are the hallmark of LMSZ interferometry. Unlike earlier approaches, our scheme does not require any approximations, allowing us to uncover important and previously unknown features of the resonance structure. We also discuss the experimental observability of our results. \end{abstract}
%%%%%%%%%%%%%

\maketitle

\textit{Introduction:-}
Driven quantum two-level systems have been extensively studied since the advent of quantum mechanics. A new wave of interest in coherent two-level dynamics has arisen recently in the context of solid state quantum devices, where rapid progress in fabrication techniques has made it possible to observe the signatures of the coherent evolution of few-level quantum systems in a solid state environment. Perhaps the most striking signatures of quantum coherence in two-level systems are the interference patterns that appear in the final state probability when the system is coherently driven through an avoided crossing repeatedly. This physical process is known as Landau-Majorana-St\"uckelberg-Zener (LMSZ) interferometry \cite{Landau_PZS32,Zener_PRSL32,Stuckelberg_HPA32,Majorana_NC32,Shevchenko_PR10}. First observed in atomic and optical systems \cite{Baruch_PRL92,Yoakum_PRL92,You_Nature11}, LMSZ interferometry has recently been demonstrated in few-level solid state systems including nitrogen vacancy centers in diamond \cite{Childress_PRA10,Fuchs_NP11,Zhang_arxiv12}, spins in quantum dots \cite{Petta_Science10,Gaudreau_NP12,Ribeiro_PRL13,Ribeiro_PRB13}, charge qubits \cite{Stehlik_PRB12}, and superconducting qubits \cite{Oliver_Science05,Berns_Nature08,Quintana_PRL13}. The abundance of applications to quantum computing, especially with superconducting qubits, has been largely responsible for the recent revival of this field. In all these manifestations, LMSZ interferometry has served as a powerful tool in measuring coherence times, mapping out the energy level diagram, and executing a target quantum evolution.

These advances in the experimental implementations of LMSZ interferometry have in turn spurred recent progress in the theoretical understanding of this phenomenon in a variety of contexts \cite{Damski_PRA06,Ashhab_PRA07,Rudner_PRL08,Berry_JPA09,Son_PRA09,Shevchenko_PR10,Ruschhaupt_NJP12,Malossi_arxiv1211,Zhang_PRA11,Satanin_PRB12,Sun_PRA12,Shevchenko_PRB12,Satanin_arxiv13}. Generally the type of periodic driving field considered most often in LMSZ experiments, simple monochromatic driving, cannot be solved exactly. Numerous approximate approaches have therefore been developed in the literature to treat this case including the rotating wave approximation \cite{Ashhab_PRA07}, perturbative Floquet theory \cite{Shirley_PR65,Creffield_PRB03,Son_PRA09}, and the adiabatic impulse model \cite{Damski_PRA06,Shevchenko_PR10}. These approaches involve either replacing the monochromatic field with a different but related driving protocol that is more amenable to an analytical treatment or expanding in the limit of small avoided crossing energy gap. Although these approaches have had some success, each method is applicable only in very specific regions of parameter space, and the results thus give a patchwork analytical understanding of LMSZ interferometry. As a consequence, much of the structure underlying the diverse interference patterns that arise in LMSZ interferometry has gone overlooked and unappreciated although it provides important insight into the relevant quantum dynamics.

In this Letter, we propose a general framework based on Floquet theory that can extract the LMSZ interference patterns without any approximations to the driving field or Hamiltonian and without solving the Schr\"odinger equation. We instead exploit the fact that the interference patterns of LMSZ experiments are intimately related to periodic evolution of the system. Constraining the evolution to be periodic leads to a condition on the Hamiltonian parameters; the parameter values which satisfy this condition give the locations of resonances in the LMSZ interference pattern. In particular, we establish that these special values of the parameters arise as the zeros of a certain infinite Floquet determinant (FD). Remarkably, these zeros trace the interference pattern exactly, and in the process provide a deeper understanding of the origin and nature of LMSZ resonances.

The advantages of this approach are twofold. First, we can exactly reconstruct the interference pattern for experimentally relevant driving fields by simply diagonalizing matrices, and for the typical case of monochromatic driving, these matrices have the additional advantage of being tridiagonal, greatly simplifying the diagonalization process. Secondly, this method distinguishes three different classes of resonances, a fundamental aspect of the resonance structure that has not been captured by other approaches. We refer to these classes as ``real", ``complex" and ``accidental" resonances. The real and complex resonances are universal features corresponding to quantum evolution that is periodic regardless of the initial conditions, i.e., the system ``resonates" at the driving frequency; these resonances are completely characterized by the zeros of the FD. On the other hand the accidental resonances correspond to non-periodic evolution, depend on the initial conditions of the system, and are characterized by both the FD zeros and a set of non-universal eigenvectors. Every resonance that occurs in LMSZ interferometry belongs to one of these three resonance classes.

\textit{Resonance classes:-} Two-level LMSZ interferometry is described by the following Hamiltonian:
\beq
H=J(t)\sigma_z+h\sigma_x,\label{ham}
\eeq
where $\sigma_z$ and $\sigma_x$ are Pauli matrices, $J(t+T){=}J(t)$ is a periodic drive field with frequency $\omega$ and period $T{=}2\pi/\omega$, and the constant $2h$ is the minimal energy gap of the avoided crossing (centered at $J{=}0$) formed between the two diabatic levels, which we label $\ket{1}$ and $\ket{2}$. In qualitative descriptions of LMSZ interferometry, an analogy with a Mach-Zehnder interferometer is sometimes drawn \cite{Oliver_Science05} in which one interprets the diabatic states as beams which separately accumulate different phases far from the anti-crossing before propagating into the anti-crossing to interfere. Although this heuristic picture explicitly assumes that the processes of interference and phase accumulation can be distinguished temporally, which is not generally the case, it still provides intuition regarding the origin of interference patterns in the final state probabilities measured after the system traverses the avoided crossing two or more times.

\begin{figure}[htb!]
  \centering
  {\includegraphics[scale=0.4]{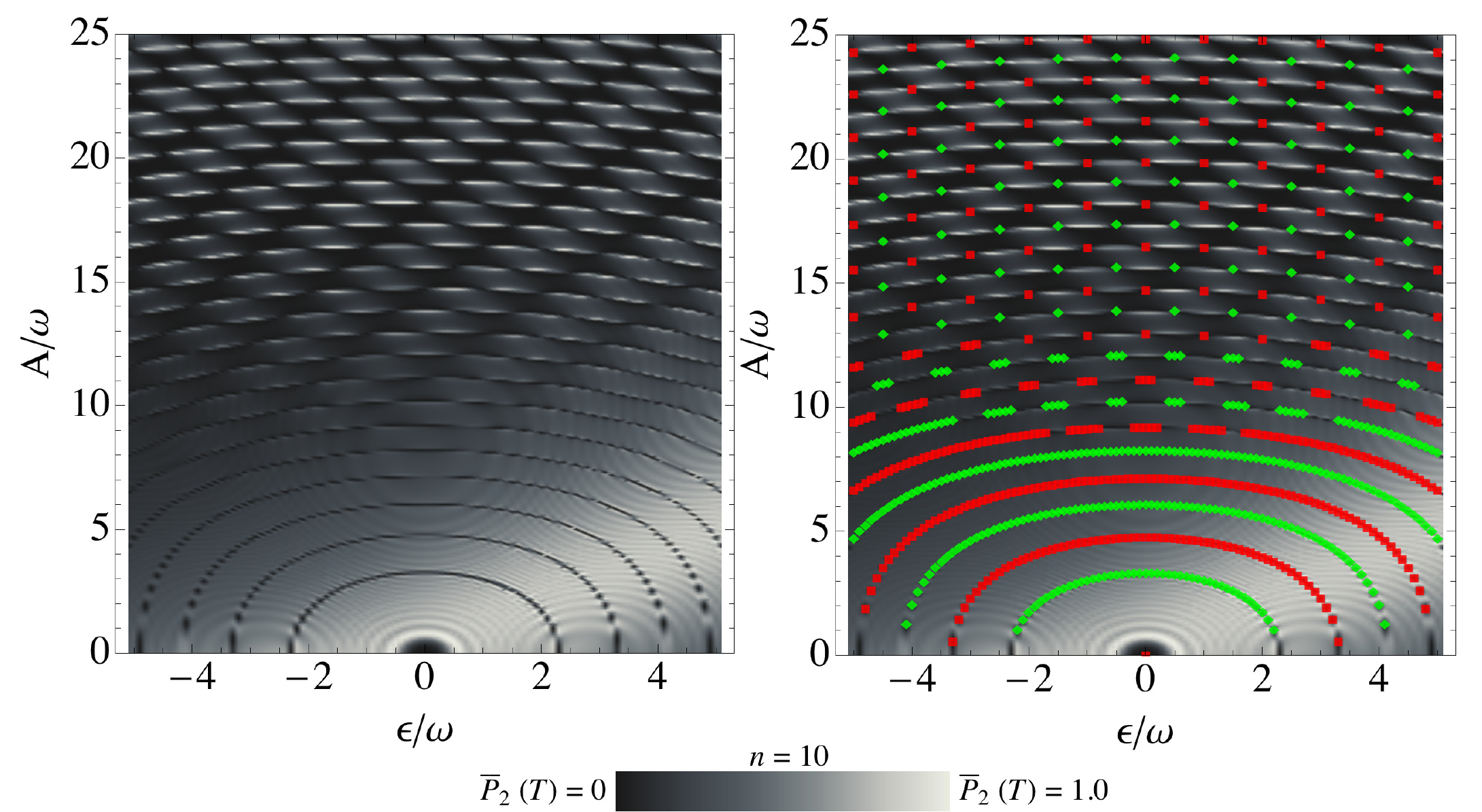}}
  \caption{(color online)Left: $\bar P_2(nT)$ as a function of drive amplitude $A$ and detuning $\epsilon$ for monochromatic driving from numerical solution of Schr\"odinger equation with $n{=}10$, $\varphi{=}{-}\pi/2$ and $h{=}5\omega$. Right: red squares and green diamonds mark real and complex roots of FD, respectively.}
  \label{densityplot}
\end{figure}

Our study of LMSZ interference patterns focuses specifically on the probability that the driving field evolves the system from state $\ket{1}$ at $t{=}0$ to state $\ket{2}$ at $t{=}nT$, corresponding to $2n$ complete traversals through the avoided crossing. We parametrize the evolution operator generated by Hamiltonian (\ref{ham}) as
\beq
U=\left(\begin{matrix}u_{11} & -u_{21}^* \cr u_{21} & u_{11}^* \end{matrix}\right),\label{generalU}
\eeq
with $u_{11}(0){=}1$, $u_{21}(0){=}0$ and $|u_{11}|^2{+}|u_{21}|^2{=}1$. The probability to be in state $\ket{2}$ after one drive period is $P_2(T){\equiv}|u_{21}(T)|^2$, while after $n$ drive periods it is
\beq
P_2(nT)=\frac{\sin^2 [n\cos^{-1}(\hbox{Re}[u_{11}(T)])]}{1-\hbox{Re}[u_{11}(T)]^2}P_2(T).
\eeq
To describe experiments in which the final state probability is averaged over many drive periods, we also consider the quantity $\bar P_2(nT){\equiv}\frac{1}{n}\sum_{m{=}1}^nP_2(mT)$. For certain values of the drive field parameters and $h$, $P_2(T)$ (and hence $P_2(nT)$ and $\bar P_2(nT)$) vanish; we refer to this point in parameter space as a resonance (also referred to as coherent destruction of tunneling \cite{Grossmann_PRL91,Stehlik_PRB12}). The collection of all such resonances forms an interference pattern. An example of such an interference pattern is shown in the left panel of Fig.~\ref{densityplot}.
%%%%%%%%%%%%%%%%%%%%%%%%%%%%%%%%%%%%%%%%%%%%%%%%%%%%%%%%%%%%%%%%%%%%%%%%%%%%%%%%%%%%%%%%
%%%%%%%%%%%%%%%%%%%%%%%%%%%%%%%%%%%%%%%%%%%%%%%%%%%%%%%%%%%%%%%%%%%%%%%%%%%%%%%%%%%%%%%%
Basic Floquet theory states that the evolution operator for a periodic Hamiltonian can be expressed as \cite{Shirley_PR65}:
\beq
U=\left(\begin{matrix}p_{11} & -p_{21}^* \cr p_{21} & p_{11}^*\end{matrix}\right)\left(\begin{matrix} e^{-iEt} & 0 \cr 0 & e^{iEt}\end{matrix}\right)\left(\begin{matrix}p_{11}^*(0) & p_{21}^*(0) \cr -p_{21}(0) & p_{11}(0)\end{matrix}\right),\label{basicFloquet}
\eeq
where $p_{11}(t+T){=}p_{11}(t)$, $p_{21}(t+T){=}p_{21}(t)$ are periodic functions with $|p_{11}|^2{+}|p_{21}|^2{=}1$, and $E$ is the quasi-energy. It is clear from this equation that the evolution is periodic only if $E{=}k\pi/T$ for some integer $k$. When $k$ is odd, the state acquires an overall minus sign at $t=T$; while this minus sign is immaterial for a system with only two levels, it can be relevant when additional levels are present. Therefore we maintain a distinction between $2\pi$ and $4\pi$-periodic evolutions arising when $k{=}0$ or $k{=}1$, respectively.  Equating (\ref{generalU}) and (\ref{basicFloquet}) gives the functions $u_{11}$ and $u_{21}$ after one full period:
\bea
u_{11}(T)&=&e^{i ET}-2i|p_{11}(0)|^2\sin(ET),\nn\\
u_{21}(T)&=&-2ip_{11}^*(0)p_{21}(0)\sin(ET).
\eea
These expressions make it clear that $P_2(T)$ can only vanish if $E{=}k\pi/T$, i.e. if the evolution is $2\pi$ or $4\pi$-periodic, or if $p_{11}^*(0)p_{21}(0){=}0$, in which cases the evolution is generally not periodic. Thus we see that the resonances naturally divide into two classes: resonances which satisfy $E{=}k\pi/T$ and correspond to periodic evolution, and resonances which obey the condition $p_{11}^*(0)p_{21}(0){=}0$ and correspond to non-periodic evolution. Resonances of the latter type will be referred to as accidental resonances. We show below that the resonances associated with periodic evolution can be further categorized into ``real" and ``complex" resonances, corresponding to real and complex roots of the FD. In the rare case where both $E{=}k\pi/T$ and $p_{11}^*(0)p_{21}(0){=}0$ hold, the resonance will be classified as real.

To our knowledge, the distinction between the different types of LMSZ resonances has not previously been identified in the literature. It is important to make this distinction because these classes have different experimental manifestations. Real and complex resonances are universal features in the sense that they are determined solely by the quasi-energy, whereas accidental resonances are non-universal since they depend on the initial data $p_{11}^*(0)$, $p_{21}(0)$. Real and complex resonances occur when the evolution of the system is periodic; the system is rotated by $2\pi$ or $4\pi$ about some axis. Accidental resonances do not generally correspond to periodic evolution. Instead, they correspond simply to a process in which the state is rotated partially about the drive axis. As shown below in the case of monochromatic driving, accidental resonances depend on the phase of the drive field and will thus be washed out in experiments which are insensitive to this phase.

\textit{Floquet determinant:-} The fact that real and complex resonances are associated with special values of the quasi-energy means that we do not need to solve the quantum evolution in order to compute the parameter values where these resonances occur. However, in order to see this, it helps to first set up a formal series solution for this evolution even though we do not need to compute this solution explicitly. For this purpose, instead of $u_{11}$ and $u_{21}$, it is convenient to work with the functions
\beq
\phi_1{=}e^{-i\int_0^tdt'J(t')}p_{21}e^{-iEt},\;\;
\phi_2{=}e^{-i\int_0^tdt'J(t')}p_{11}^*e^{iEt},
\eeq
which are independent solutions of the following second-order Schr\"odinger equation:
\beq
\ddot\phi+2iJ\dot\phi+h^2\phi=0.\label{scheq}
\eeq
The motivation for introducing $\phi_1$ and $\phi_2$ is that since $J$, $p_{11}$, and $p_{21}$ are each periodic functions, $\phi_1$ and $\phi_2$ admit Fourier series expansions:
\beq
\phi_1{=}\sum_{m=-\infty}^\infty a_m e^{i(m+\alpha_+)\omega t},\;\;
\phi_2{=}\sum_{m=-\infty}^\infty b_m e^{i(m+\alpha_-)\omega t},
\eeq
where $\alpha_\pm{=}(\mp E{-}\epsilon)/\omega$, and we have allowed for a possible overall additive constant $\epsilon$ in $J(t)$ such that $\int_0^Tdt'J(t'){=}\epsilon T$. We may think of $\epsilon$ as a drive field detuning parameter. Plugging these expansions into the Schr\"odinger equation (\ref{scheq}) along with a similar expansion for the drive field, $J{=}\sum_m j_m e^{im\omega t}$, we obtain a recursion relation for the Fourier coefficients:
\beq
2\omega\sum_\ell j_{m-\ell}(\ell+\alpha_+)a_\ell+[\omega^2(m+\alpha_+)^2-h^2]a_m=0,\label{genrecur}
\eeq
and similarly for $b_m$ with $\alpha_+{\to}\alpha_-$. We can organize the coefficients in this recursion relation into an infinite matrix $M_+$ which acts on the infinite vector, $a{\equiv}(\ldots,a_{m-1},a_m,a_{m+1},\ldots)$, recasting (\ref{genrecur}) as the condition that $a$ is a null vector of $M_+$: $M_+ a{=}0$. Similarly, $b{\equiv}(\ldots,b_{m-1},b_m,b_{m+1},\ldots)$ is a null vector of a matrix $M_-$. For generic Hamiltonian parameters, the null spaces of $M_+$ and $M_-$ must both be non-empty for some value of $E$ since the Schr\"odinger equation always has a solution. Since a non-empty null space implies the vanishing of the determinant, if we impose the periodicity condition $E{=}k\pi/T$, we obtain a general formula for the locations of real and complex resonances:
\beq
\det M_k=0,\label{floquetdet}
\eeq
where $M_k{\equiv}M_+({\alpha_+{=}{-}k/2{-}\epsilon/\omega})$. We refer to $\det M_k$ as a Floquet determinant. Eq.~(\ref{floquetdet}) is one of the main results of this paper. Given a periodic drive field $J(t)$ with frequency $\omega$ and detuning $\epsilon$, we can systematically construct the matrix $M_k$, which depends only on drive parameters and the energy gap $h$. A set of parameters which solves Eq.~(\ref{floquetdet}), i.e., a root of the FD, corresponds to the location of a resonance in the LMSZ interference pattern, with real and complex roots giving real and complex resonances respectively. Physical solutions are of course given by real parameter values; however the real parts of complex roots with small imaginary parts still give the locations of approximate resonances. Hence, complex resonances are approximate resonances, whereas real resonances are exact. Note that it suffices to solve (\ref{floquetdet}) only for the cases $k{=}0$ and $k{=}1$ in order to obtain the full set of resonances in the interference pattern. Other choices of $k$ are equivalent to these since the recursion relation (\ref{genrecur}) is invariant under a shift of $k$ by an even integer. Also note that we do not need to separately solve the condition (\ref{floquetdet}) with $\alpha_+$ replaced by $\alpha_-$ since $M_+{=}M_-{=}M_k$ when $E{=}k\pi/T$. This degeneracy translates to a twofold degeneracy in the solutions of Eq.~(\ref{floquetdet}).

%%%%%%%%%%%%%%%%%%%%%%%%%%%%%%%%%%%%%%%%%%%%%%%%%%%%%%%%%%%%%%%%%%%%%%%%%%%%%%%%%%%%%%%%%?
\begin{center}
\begin{figure}[htb!]
  \includegraphics[scale=0.42]{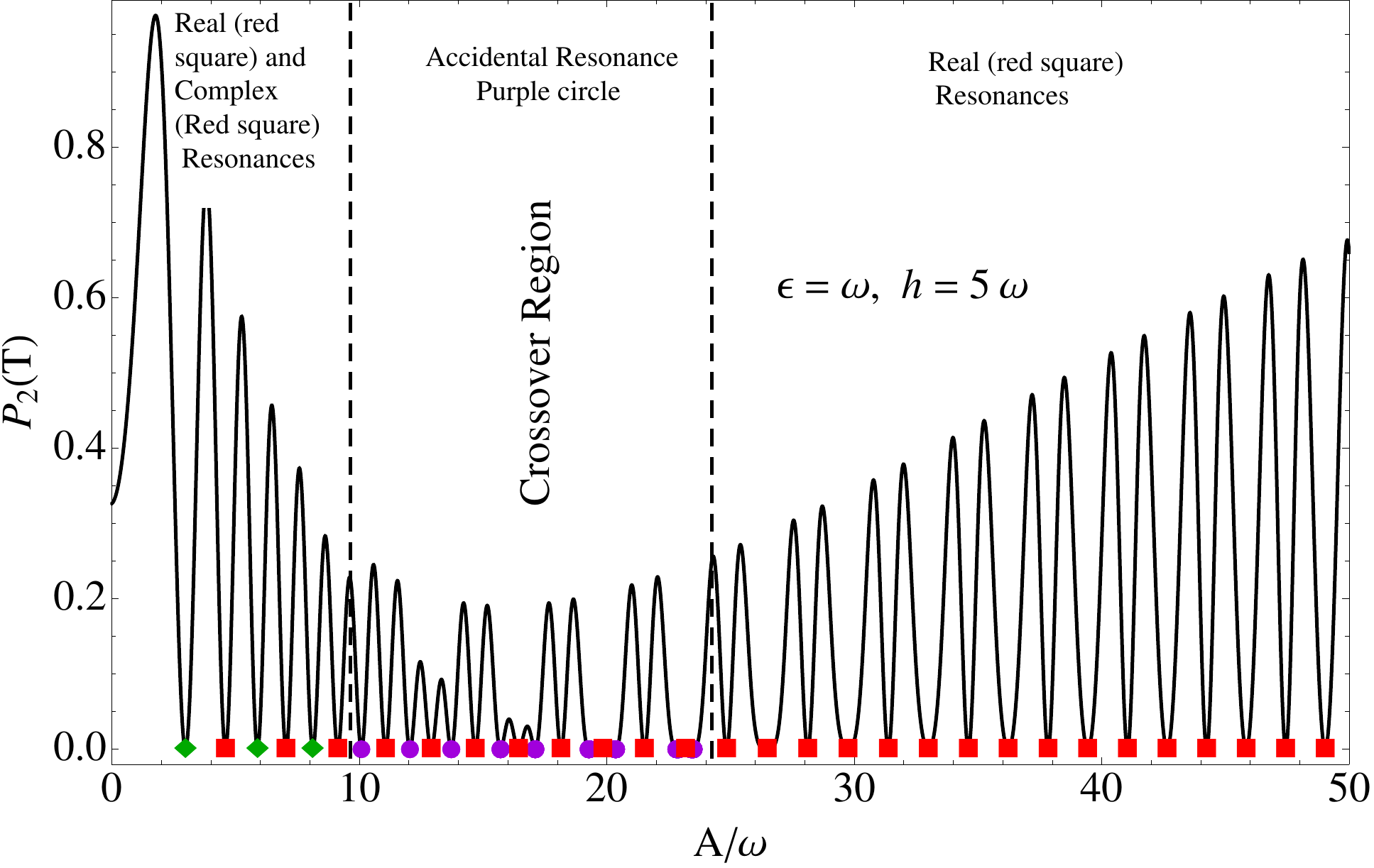}
  \caption{$P_2(T)$ from numerical solution of Schr\"odinger equation with $h{=}5\omega$, $\epsilon{=}\omega$, $\varphi{=}{-}\pi/2$. Red squares, green diamonds and purple circles mark real, complex and accidental resonances obtained from the FD, respectively.}
\label{p2a}
\end{figure}
\end{center}
%%%%%%%%%%%%%%%%%%%%%%%%%%%%%%%%%%%%%%%%%%%%%%%%%%%%%%%%%%%%%%%%%%%%%%%%%%%%%%%%%%%%%%%%%
\vspace{-1.0cm}
A further crucial point pertains to how one should solve Eq.~(\ref{floquetdet}). Notice that $M_k$ is a linear function of both the driving field amplitude $A$ (since the $j_m$ are linear in $A$), and $h^2$. This means that if we want to fix all other parameters and solve for either $A$ or $h^2$, we may do so by re-interpreting $M_k a{=}0$ as either an ordinary (in the case of $h^2$) eigenvalue problem or as a generalized (in the case of $A$) eigenvalue problem. For instance, if we wish to fix all driving field parameters and solve for the values of $h$ that correspond to real or complex resonances, we should rewrite the null space condition $M_k a{=}0$ as $M_k^{h{=}0} a=h^2a$, so that the resonances are given by the eigenvalues of $M_k^{h{=}0}$. In practice, we must truncate the infinite matrix $M_k$ in order to solve this eigenvalue problem, and the issue of convergence with respect to truncation size becomes important. We comment on this further in the specific case of monochromatic driving presented below.

\textit{Monochromatic driving:-} We illustrate the general analysis given above with the example of monochromatic driving, which is a common choice in LMSZ interferometry experiments (see e.g., Ref.~\cite{Stehlik_PRB12}):
\beq
J(t)=\epsilon+A\sin(\omega t+\varphi).\label{monodrive}
\eeq
The matrix $M_k$ for this driving field is given in the appendix. As described above, we fix all parameters except for $A$ and solve $\det M_k{=}0$, treating $A$ as the generalized eigenvalue to be computed. As a representative example we fix $h{=}5\omega$, scan $\epsilon$ and solve for $A$ for both $k{=}0$ and $k{=}1$. Convergence with respect to the truncation size $N$ of $M_k$ is very rapid, with full convergence of the eigenvalues $A{<}A_{max}$ occurring roughly for $N{\ge} 4A_{max}$. Eigenvalues larger than $A_{max}$ gradually become inaccurate due to the truncation. In Fig.~\ref{densityplot}, we overlay the resulting roots on the interference pattern obtained by solving the Schr\"odinger equation numerically, directly demonstrating that the FD roots accurately trace the contours of resonances for all values of $\epsilon$. In Fig.~\ref{p2a}, we show a slice of the interference pattern with $\epsilon{=}\omega$. Remarkably, all the resonance points evident in the numerical curve are exactly captured by the characteristic values of $A$. Moreover, it is clear from the figure that for smaller driving amplitudes, the resonances alternate between real and complex (corresponding to $2\pi$ and $4\pi$ periodicity respectively), while for larger amplitudes, all the resonances are real. We admit complex resonances with imaginary part less than $10^{-2}$. The fact that real resonances are singled out at larger amplitudes can be understood from the fact that in this regime, the energy gap $h$ can be neglected relative to the driving amplitude, and the quasi-energy is simply given by the detuning: $E{=}\pm\epsilon$. For the example shown in Fig.~\ref{p2a}, $\epsilon$ is an integer, so that only resonances corresponding to integer values of the quasi-energy, namely the $2\pi$-periodic resonances, appear. This argument also reveals that no real or complex resonances occur at large amplitude when $2\epsilon$ is not an integer, a fact which is evident in the right panel of Fig.~\ref{densityplot}. The alternating behavior at lower amplitude can be traced to the alternating integer and half integer eigenvalues of the harmonic oscillator, which coincides with the monochromatic driving Hamiltonian in the limit $h^2\gg A$. Fig.~\ref{p2a} shows a slice of the resonance pattern after one traversal through the avoided crossing; in the appendix we show that resonances can either become sharper or disappear altogether when the final state probability is averaged over many periods. This behavior is particularly relevant for the experimental observability of LMSZ resonances.
%%%%%%%%%%%%%%%%%%%%%%%%%%%%%%%%%%%%%%%%%%%%%%%%%%%%%%%%%%%%%%%%%%%%%%%%%%%%%%%%%%%%%%%%%
\begin{center}
\begin{figure}[htb!]
\includegraphics[scale=0.42]{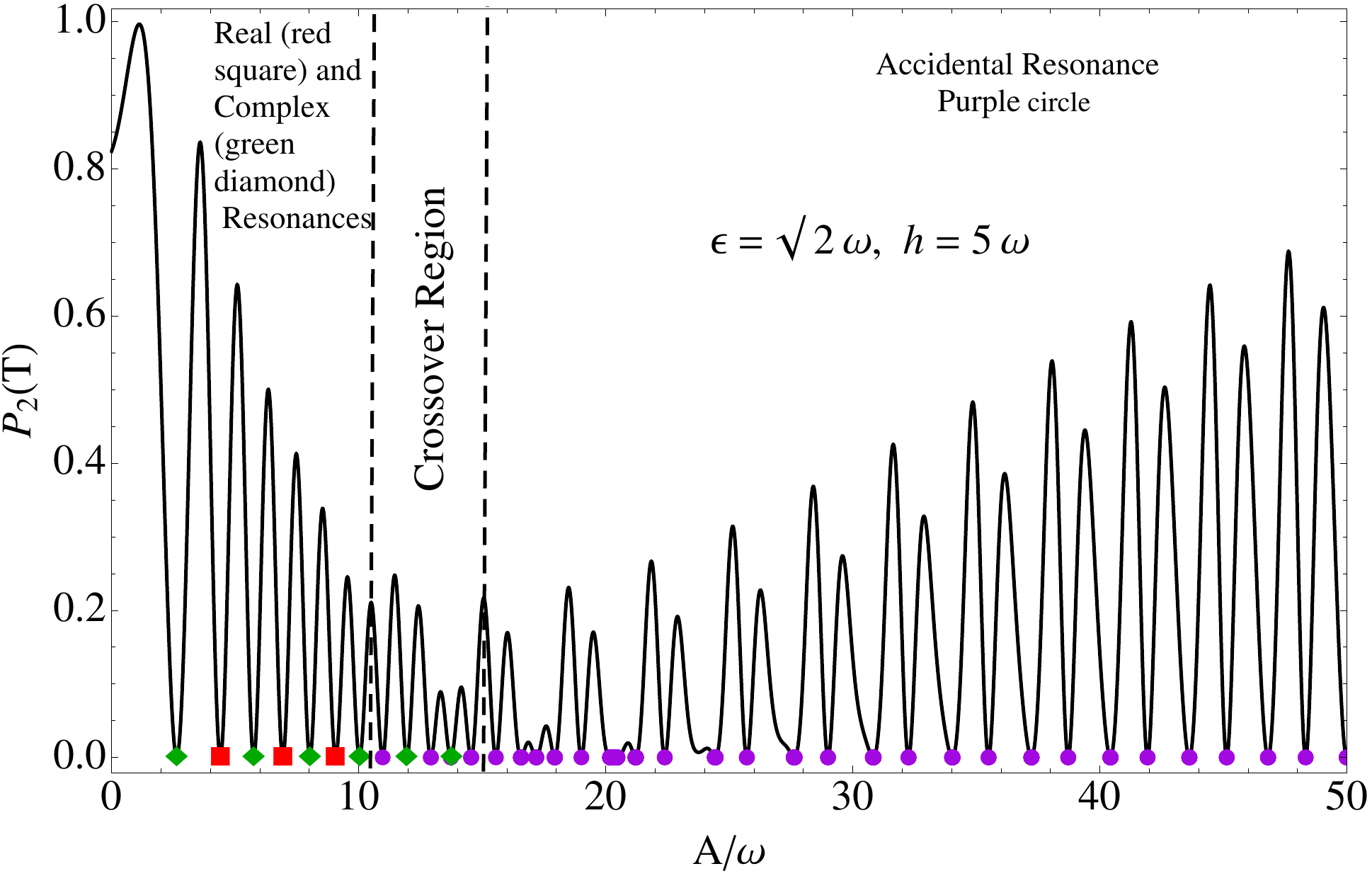}
\caption{$P_2(T)$ from numerical solution of Schr\"odinger equation with $h{=}5\omega$, $\epsilon{=}\sqrt{2}\omega$, $\varphi{=}{-}\pi/2$. Red squares, green diamonds and purple circles mark real, complex and accidental resonances obtained from the FD, respectively.}
\label{ar}
\end{figure}
\end{center}
%%%%%%%%%%%%%%%%%%%%%%%%%%%%%%%%%%%%%%%%%%%%%%%%%%%%%%%%%%%%%%%%%%%%%%%%%%%%%%%%%%%%%%%%%
\vspace{-1.5cm}
%%%%%%%%%%%%%%%%%%%%%%%%%%%%%%%%%%%%%%%%%%%%%%%%%%%%%%%%%%%%%%%%%%%%%%%%%%%%%%%%%%%%%%%%%
\begin{center}
\begin{figure}[htb!]
\includegraphics[scale=0.41]{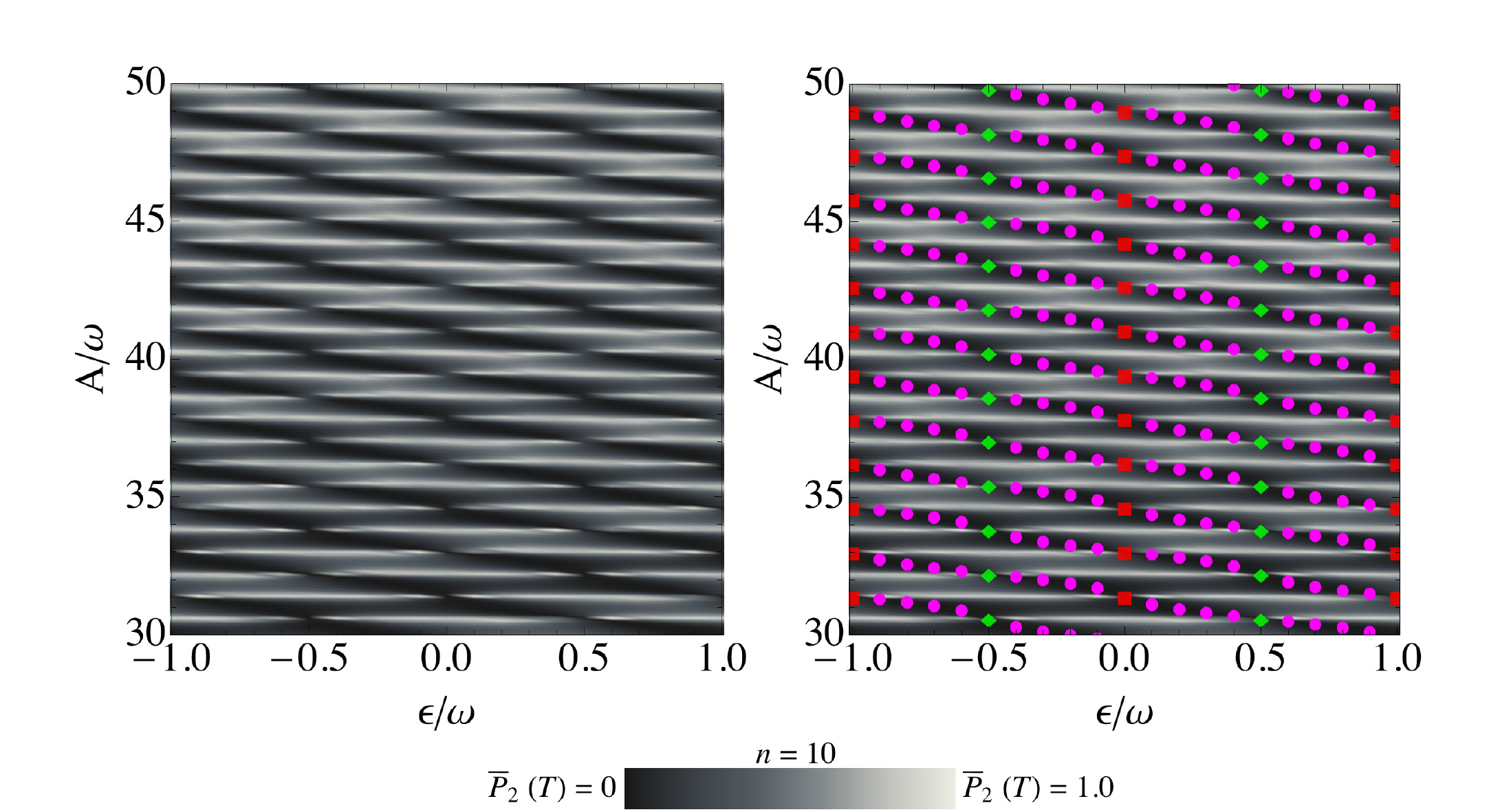}
\caption{$P_2(T)$ as a function of $A$ and $\epsilon$ for the parameters of Fig.~\ref{densityplot}. Real (red square) and complex (green diamond) resonances only occur for $2\epsilon{=}n\omega$, while accidental (purple circles) resonances occur for intermediate values of $2\epsilon$.}
\label{arden}
\end{figure}
\end{center}
%%%%%%%%%%%%%%%%%%%%%%%%%%%%%%%%%%%%%%%%%%%%%%%%%%%%%%%%%%%%%%%%%%%%%%%%%%%%%%%%%%%%%%%%%
\vspace{-1.0cm}

\textit{Accidental resonances:-} In addition to real and complex resonances, the FD method can also capture the less universal yet interesting class of accidental resonances (AR). The defining criterion for ARs, $p_{11}^*(0)p_{21}(0){=}0$, can be re-expressed as the condition that the sum of the components of one of the null vectors, $a$ or $b$, vanishes: $\left(\sum_\ell a_\ell\right)\left(\sum_m b_m\right){=}0$. Unlike the cases of real and complex resonances, the quasi-energy associated with an AR can have any value. However, since the quasi-energy is only defined modulo integer multiples of $\omega$, all the ARs can be obtained by scanning over $0{\le }E{<}\omega$, solving the FD in each case, and keeping those resonances which satisfy $\sum_m a_m{=}0$ or $\sum_m b_m{=}0$.
Fig.~\ref{ar} shows the ARs that result from this method along with FD results for real and complex resonances for a particular set of parameters, illustrating that all the resonances are captured by the FD method. It is typically the case that the conditions $\sum_m a_m{=}0$ and $\sum_m b_m{=}0$ are satisfied only approximately, in which case it is necessary to introduce a small upper cutoff (of the order $|\sum_m a_m|{<}10^{-2}$) on the sums of $a_m$ and $b_m$ in order to compute approximate ARs. In  Fig.~\ref{ar}, we note that for smaller values of $A$, resonances are either real or complex, while for larger values of $A$ all resonances are accidental, implying a crossover region where real or complex resonances are converted to accidental resonances. Fig.~\ref{arden} shows the same density plot of Fig.~\ref{densityplot} but in a region of larger drive amplitude. The figure reveals that this conversion process happens for any non-integer value of $2\epsilon$, while for integer values, the resonances remain real or complex beyond the crossover region.

\textit{Conclusion:-} We have presented a general framework for calculating and characterizing the resonance structure of LMSZ interferometry from properties of an infinite Floquet determinant. This framework uncovers a natural classification of LMSZ resonances into three basic categories distinguished by whether these resonances are exact or approximate and by whether they correspond to periodic or non-periodic quantum evolution. Our approach applies to all parameter regimes, giving a unified picture of LMSZ interferometry.

\textit{Acknowledgments:-} We thank Lev Bishop and Benjamin Fregoso for helpful discussions. This work is supported by LPS-NSA, IARPA and JQI-NSF-PFC.

\bigskip
\centerline{\bf{Appendix}}

In this appendix, we consider further details of the Floquet Determinant analysis applied to the case of monochromatic driving. We focus particularly on the visibility of LMSZ resonances in experiments where the final state probability is averaged over many drive periods. We also consider the sensitivity of the resonances to the phase of the driving field. For monochromatic driving, the external field has the form $J(t){=}\epsilon+A\sin(\omega t+\varphi)$, and the Floquet matrix $M_k$ is
\bea
M_k&=&\left(\begin{matrix} . & . & . & . & . \cr . & v_{m-1} & w_{m-1} & 0 & . \cr . & w_{m-2}^* & v_m & w_m & . \cr . & 0 & w_{m-1}^* & v_{m+1} & . \cr . & . & . & . & .\end{matrix}\right),\nn\\
w_m&\equiv& iA\omega e^{-i\varphi}(m+\alpha_k+1),\nn\\
v_m&\equiv& \omega^2(m+\alpha_k)^2+2\epsilon\omega(m+\alpha_k)-h^2,
\eea
with $\alpha_k{\equiv}{-}k/2{-}\epsilon/\omega$.

\begin{center}
\begin{figure}[htb!]
  \includegraphics[scale=0.41]{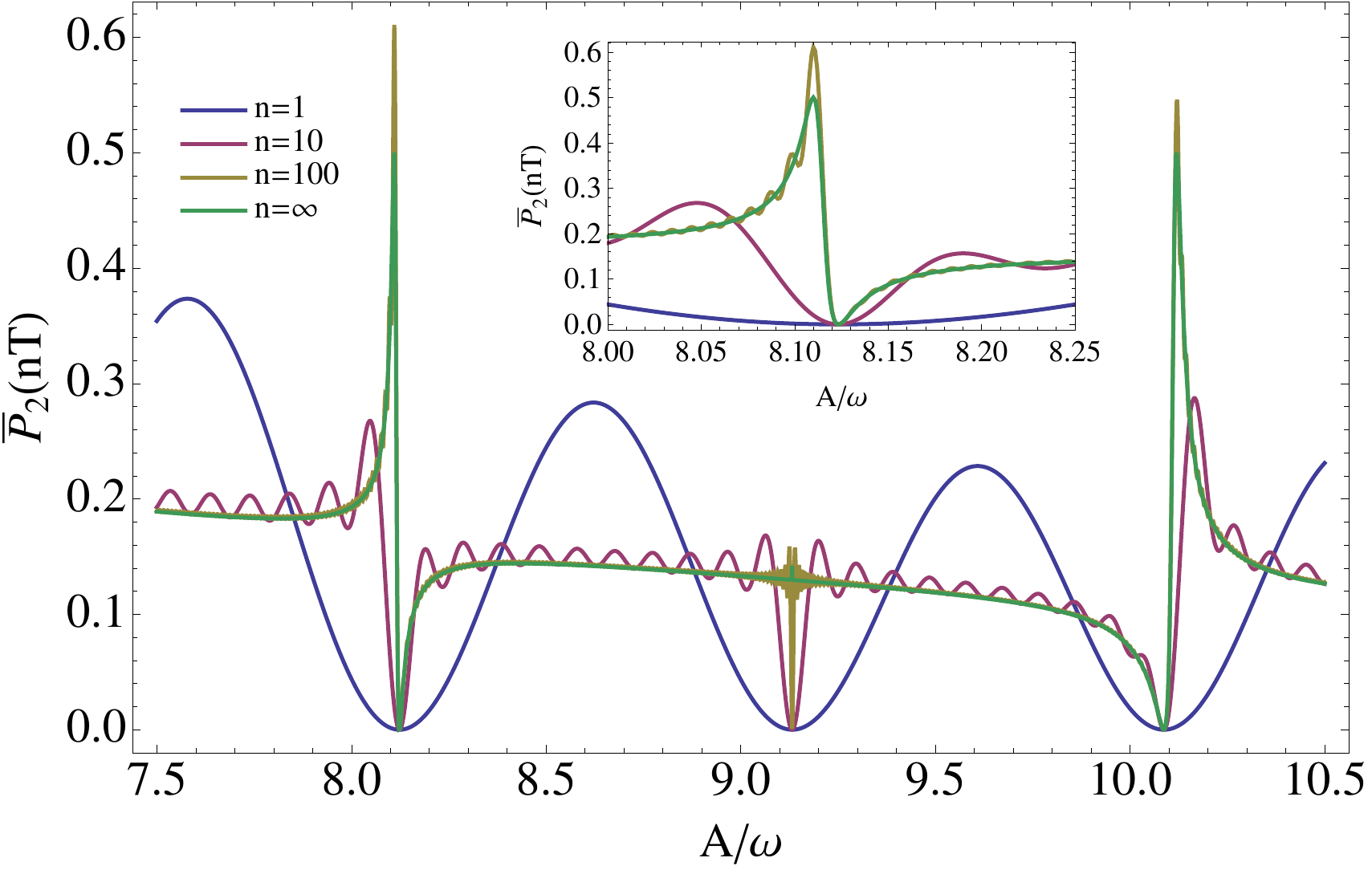}
    \caption{$\bar{P}_2(nT)$ from numerical solution of Schr\"odinger equation with $h{=}5\omega$, $\epsilon{=}\omega$, $\varphi{=}{-}\pi/2$. Different curves correspond to $2n$ complete traversals through the avoided crossing.}
\label{p2an}
\end{figure}
\end{center}

\begin{center}
\begin{figure}[htb!]
  \subfigure[$\bar{P}_2(nT)$ vs. $A/ \omega$ for $\varphi{=}{-}\pi/2$, $\epsilon{=}1.7\omega$, $h{=}5\omega$ averaged over $n{=}1000$ periods.]{\includegraphics[scale=0.35]{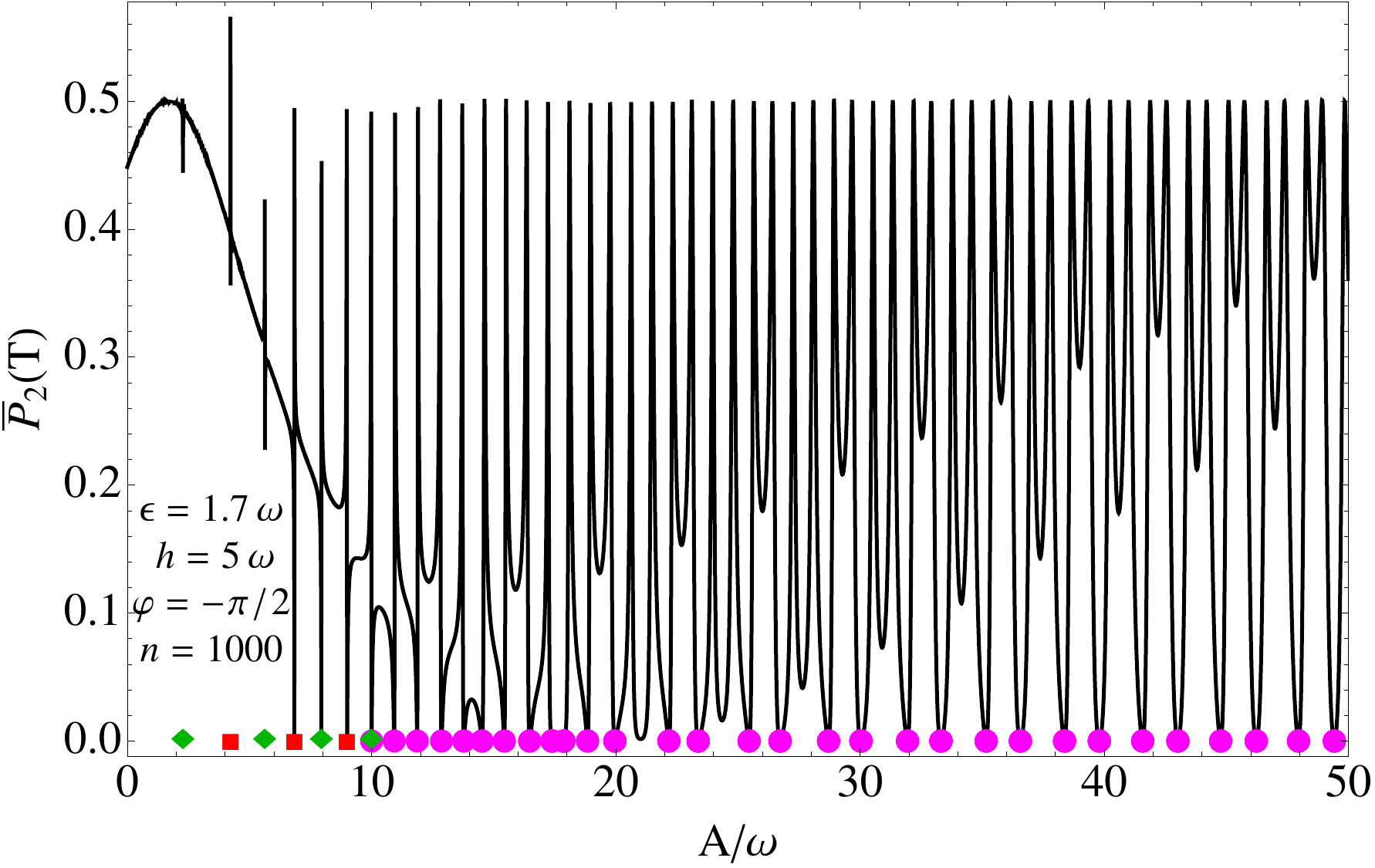}}
  \subfigure[$\bar{P}_2(nT)$ vs. $A/ \omega$ for $\varphi{=}{-}\pi/2$, $\epsilon{=}\omega$, $h{=}5\omega$ averaged over $n{=}1000$ periods.]{\includegraphics[scale=0.35]{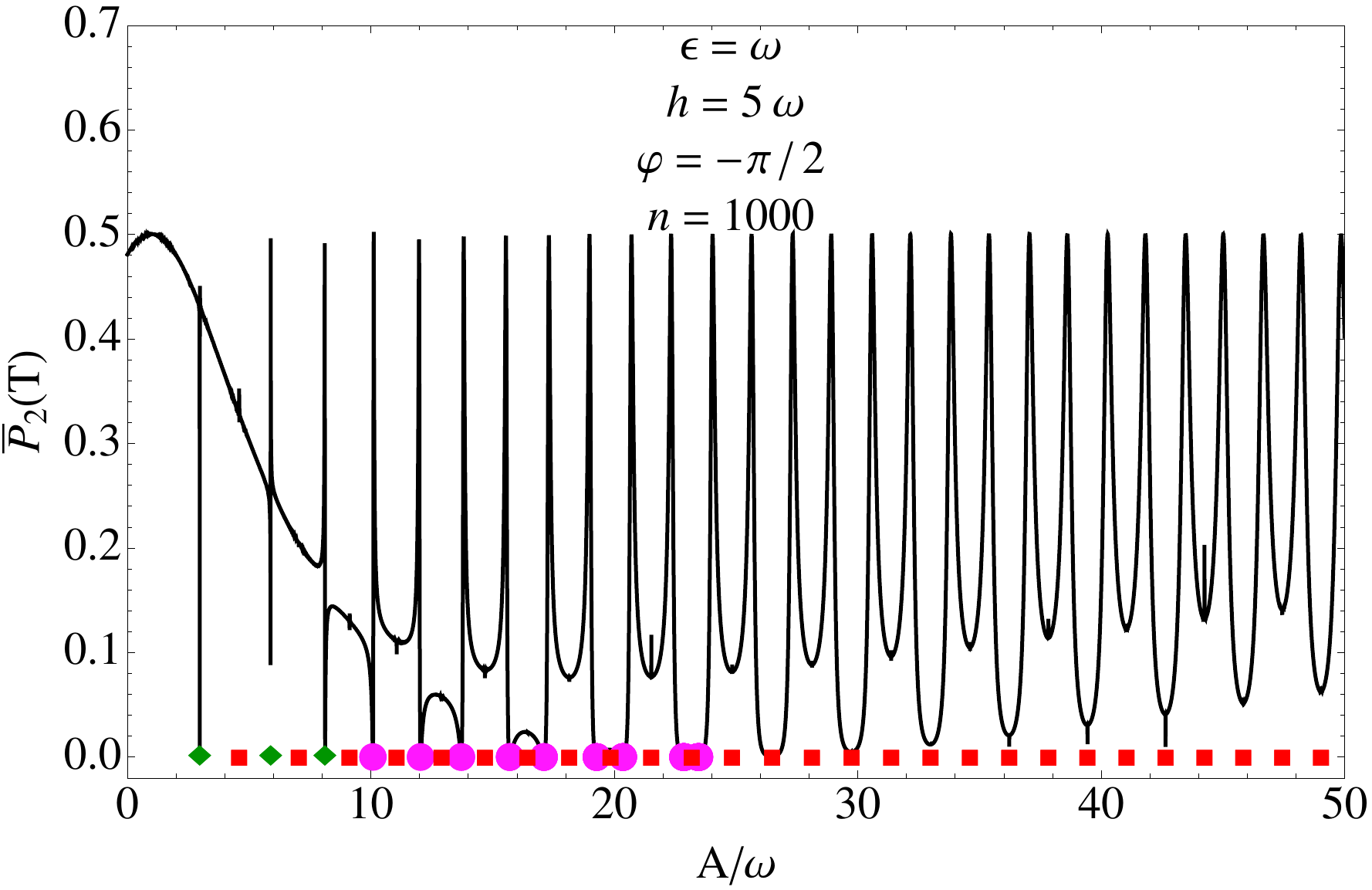}}
  \caption{Plots showing evolution corresponding to averaging over $n{=}1000$ periods. Red square and green diamonds mark real and complex roots of Floquet determinant. Purple dots correspond to the accidental resonances.}
\label{n1000}
\end{figure}
\end{center}

{\it Experimental observability:-}
In this section, we elaborate on the experimental observability of the LMSZ resonance structure. In particular, we analyze the visibility of individual resonances as a function of the number of traversals $n$ through the avoided crossing. We focus on the case where the final state probability is averaged over several periods, i.e., we consider $\bar P_2(nT)$. As demonstrated in Fig.~\ref{p2an}, the resonances generally become sharper as $n$ is increased. Although some of the resonances disappear in the limit of large $n$, most are robust and visible as $n$ becomes large.
%%%%%%%%%%%%%%%%%%%%%%%%%%%%%%%%%%%%%%%%%%%%%%%%%%%%%%%%%%%%%%%%%%%%%%%%%%%%%%%%%%%%%%%%%?
%%%%%%%%%%%%%%%%%%%%%%%%%%%%%%%%%%%%%%%%%%%%%%%%%%%%%%%%%%%%%%%%%%%%%%%%%%%%%%%%%%%%%%%%%
In Fig.~\ref{n1000}, we show how the resonance structure changes for $n{=}1000$ for periodic and non-periodic evolution. In the top panel of Fig.~\ref{n1000}, we see that the accidental resonances are very robust against the averaging over large $n$. The lower panel shows that the real and complex resonances become sharper and narrower, but retain the signatures of a resonance in the form of kinks. In this plot too we see that the accidental resonances are quite robust against the averaging process.
%%%%%%%%%%%%%%%%%%%%%%%%%%%%%%%%%%%%%%%%%%%%%%%%%%%%%%%%%%%%%%%%%%%%%%%%%%%%%%%%%%%%%%%%%?
%%%%%%%%%%%%%%%%%%%%%%%%%%%%%%%%%%%%%%%%%%%%%%%%%%%%%%%%%%%%%%%%%%%%%%%%%%%%%%%%%%%%%%%%%

\begin{center}
\begin{figure}
  \includegraphics[scale=0.41]{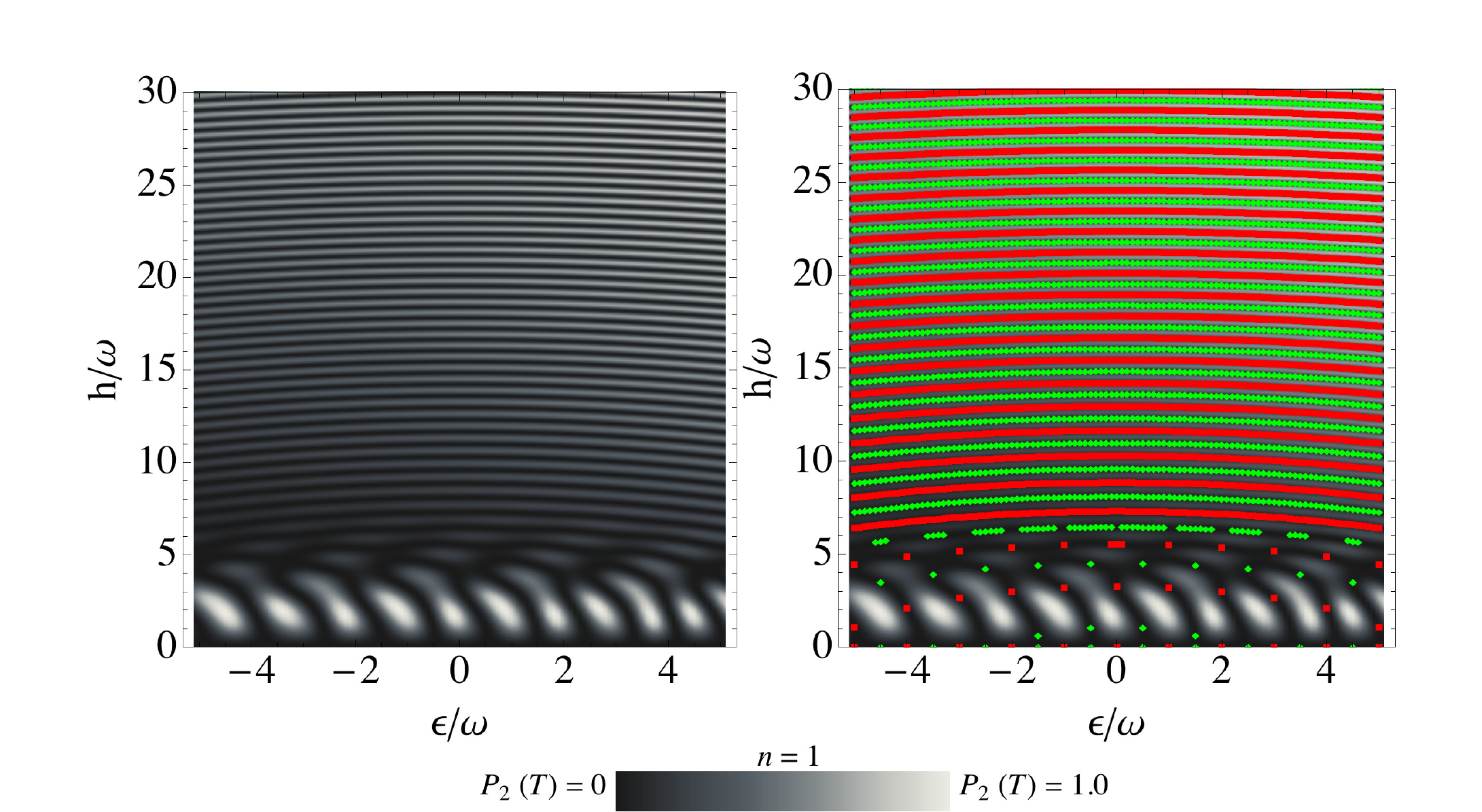}
  %\subfigure[$P_2(nT)$ vs. $A/ \omega$ for $n=50$.]{\includegraphics[scale=0.48]{p2an50.pdf}}\quad\quad
  \caption{Left Panel: Density plot of $P_2(T)$ from numerical solution of Schr\"odinger equation with $A{=}16\omega$, $\varphi{=}{-}\pi/2$. Right Panel: Red square and green diamonds mark real and complex roots of Floquet determinant.}
\label{fulldensity}
\end{figure}
\end{center}

{\it Detuning ($\epsilon$) vs. gap ($h$) density plots:-}
In the main text we presented the density plots in terms of the variables $\epsilon$ and $A$ with the minimum gap between the levels fixed at some value of $2h$. In this section we trace the interference contours by collecting the roots of the Floquet Determinant (FD) corresponding to $2h$ for different values of $\epsilon$ (see  Fig.~\ref{fulldensity}).
%%%%%%%%%%%%%%%%%%%%%%%%%%%%%%%%%%%%%%%%%%%%%%%%%%%%%%%%%%%%%%%%%%%%%%%%%%%%%%%%%%%%%%%%%
%%%%%%%%%%%%%%%%%%%%%%%%%%%%%%%%%%%%%%%%%%%%%%%%%%%%%%%%%%%%%%%%%%%%%%%%%%%%%%%%%%%%%%%%%

\begin{center}
\begin{figure}
  \includegraphics[scale=0.38]{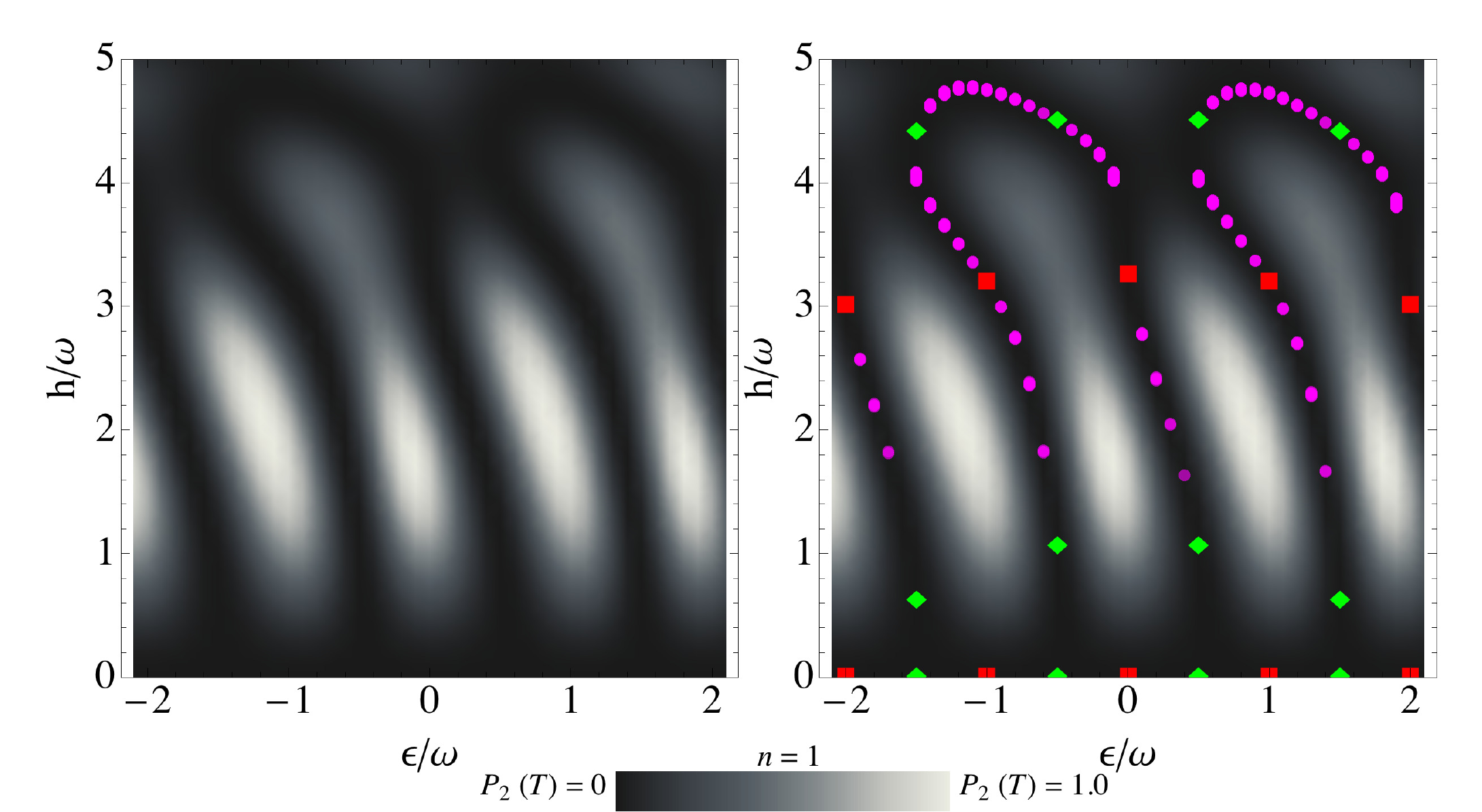}
  %\subfigure[$P_2(nT)$ vs. $A/ \omega$ for $n=50$.]{\includegraphics[scale=0.48]{p2an50.pdf}}\quad\quad
  \caption{Density plot of $P_2(T)$ from numerical solution of Schr\"odinger equation with $A{=}16\omega$, $\varphi{=}{-}\pi/2$. Red square and green diamonds mark real and complex roots of Floquet determinant. Purple dots correspond to the accidental resonances.}
\label{accidentaldensity}
\end{figure}
\end{center}

 In Fig.~\ref{fulldensity} we see that the crossover region pointed out in the main text occurs approximately for $h\lesssim\sqrt{A}$. For $h\gg\sqrt{A}$ we see that the evolution is harmonic oscillator like.  In this regime, the real and complex resonances alternate in correspondence with the integer and half integer eigenvalues and eigenstates of the simple harmonic oscillator. The evolution becomes interesting in the region $h<\sqrt{A}$, where all three type of resonances coexist as shown in Fig.~\ref{accidentaldensity}.
%%%%%%%%%%%%%%%%%%%%%%%%%%%%%%%%%%%%%%%%%%%%%%%%%%%%%%%%%%%%%%%%%%%%%%%%%%%%%%%%%%%%%%%%%
%%%%%%%%%%%%%%%%%%%%%%%%%%%%%%%%%%%%%%%%%%%%%%%%%%%%%%%%%%%%%%%%%%%%%%%%%%%%%%%%%%%%%%%%%
%%%%%%%%%%%%%%%%%%%%%%%%%%%%%%%%%%%%%%%%%%%%%%%%%%%%%%%%%%%%%%%%%%%%%%%%%%%%%%%%%%%%%%%%%
\begin{center}
\begin{figure}
  \subfigure[$P_2(T)$ vs. $A/ \omega$ for $\varphi=-\pi/2$.]{\includegraphics[scale=0.55]{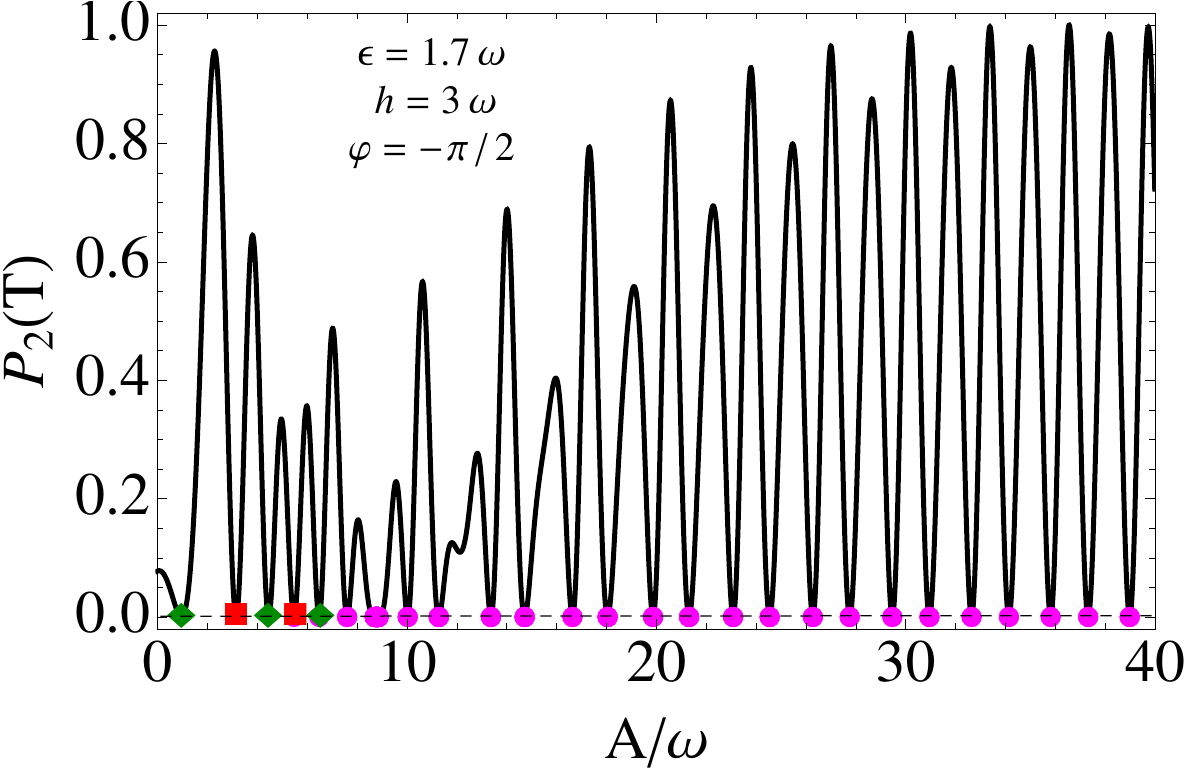}}
  \subfigure[$P_2(T)$ vs. $A/ \omega$ for  $\varphi=0$.]{\includegraphics[scale=0.55]{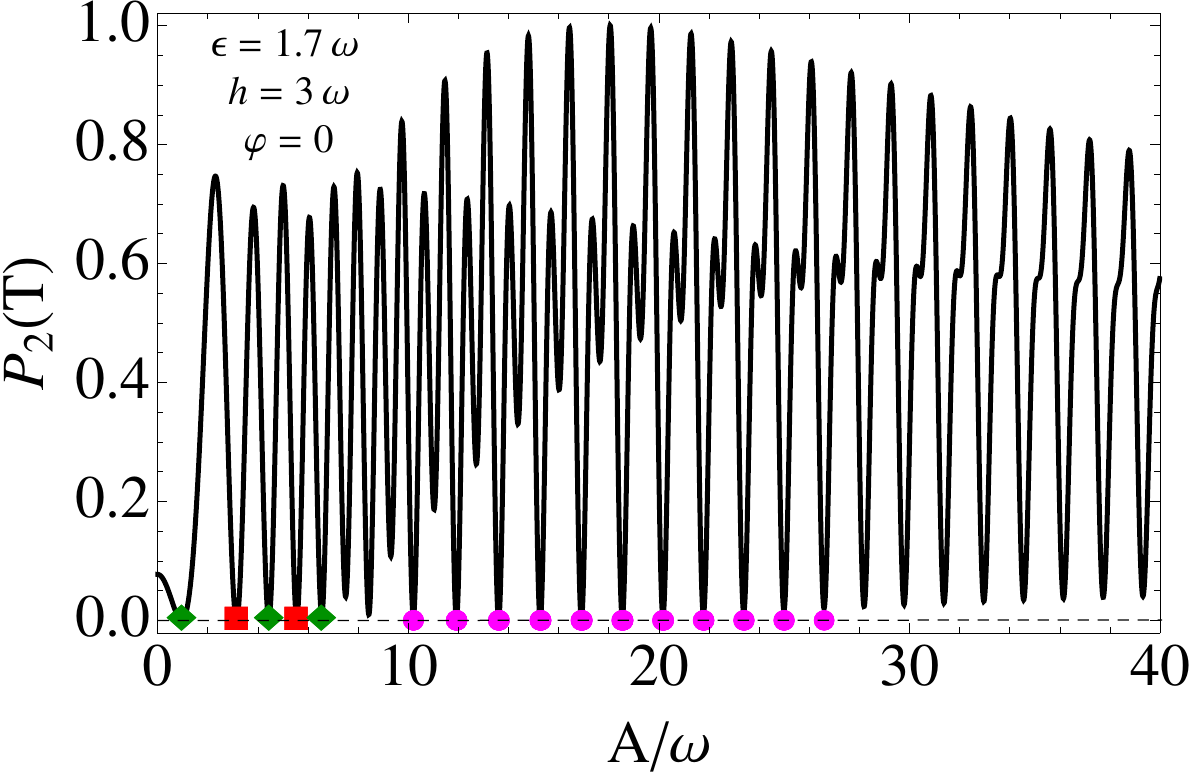}}
  \caption{Plots contrasting evolution corresponding to different driving phases for $\epsilon{=}1.7\omega$, $h{=}3\omega$. Red square and green diamonds mark real and complex roots of Floquet determinant. Purple dots correspond to the accidental resonances with cutoff $|\sum_m a_m|<10^{-2}$ for $\varphi=-\pi/2$ and and $|\sum_m a_m|<10^{-1}$ for  $\varphi=0$.}
\label{initialcond}
\end{figure}
\end{center}
%%%%%%%%%%%%%%%%%%%%%%%%%%%%%%%%%%%%%%%%%%%%%%%%%%%%%%%%%%%%%%%%%%%%%%%%%%%%%%%%%%%%%%%%%
{\it Dependence on driving phase:-} In this section we highlight how the resonance structure depends on the phase $\varphi$ of the monochromatic driving field. In Fig.~\ref{initialcond}, we compare $P_2(T)$ vs. $A/ \omega$ plotted for $\varphi{=}{-}\pi/2,\, 0$. We note that the `universal' real and complex resonances are insensitive to $\varphi$. On the other hand, the `non-universal' accidental resonance structure is sensitive to the phase $\varphi$ as shown in Fig.~\ref{initialcond}. Accidental resonances are absent within the cutoff of $|\sum_m a_m|<10^{-2}$ for the case of $\varphi=0$. We relax the cutoff to $|\sum_m a_m|<10^{-1}$ to capture approximate ARs for $\varphi=0$.

\end{document}